\documentclass[twocolumn,prl,aps,superscriptaddress,showpacs]{revtex4-1}

 \usepackage{mathptmx}
\usepackage{dcolumn}
\usepackage{amsmath,amssymb,amsfonts}
\usepackage{dsfont}
\usepackage{bm}
\usepackage{color}
\usepackage{latexsym}
\usepackage{epstopdf}
\usepackage[english]{babel}
\usepackage{enumerate}

\usepackage[pdftex]{graphicx}

\usepackage{natbib}
\usepackage{epstopdf}
\usepackage{appendix}

\definecolor{mygrey}{gray}{0.35}
\definecolor{myblue}{rgb}{0.2,0.2,0.8}
\definecolor{myzard}{cmyk}{0,0,0.05,0}
\definecolor{mywhite}{rgb}{1,1,1}
\definecolor{mywhite}{rgb}{1,1,1}
\definecolor{myred}{rgb}{1,0.,0.3}

\usepackage[colorlinks=true,citecolor=myblue,linkcolor=myred]{hyperref}

\newcommand{\bra}[1]{\left\langle #1\right|}
\newcommand{\ket}[1]{\left| #1\right\rangle}

\newcommand\nn{\mathbf{n}}

\newcommand\kk{\mathbf{k}}

\newcommand\RR{\mathbf{R}}
\newcommand\vv{\mathbf{v}}

\newcommand\hc{\mathrm{H.c.}}

\newcommand\intt{\mathrm{int}}

\begin{document}

\title{Engineering and harnessing giant atoms in high-dimensional baths: a cold atoms' implementation}
 \author{A.~Gonz\'{a}lez-Tudela}
 \email{a.gonzalez.tudela@csic.es}
 \affiliation{Instituto de F\'isica Fundamental IFF-CSIC, Calle Serrano 113b, Madrid 28006, Spain.}
  \author{C.~S\'{a}nchez~Mu\~noz}
 \affiliation{Clarendon Laboratory, University of Oxford, Oxford OX13PU, UK}
   \author{J. I.~Cirac}
  \affiliation{Max-Planck-Institut f\"{u}r Quantenoptik Hans-Kopfermann-Str. 1. 85748 Garching, Germany }

\begin{abstract}
Emitters coupled simultaneously to distant positions of a photonic bath, the so-called giant atoms, represent a new paradigm in quantum optics. When coupled to one-dimensional baths, as recently implemented with transmission lines or SAW waveguides, they lead to striking effects such as chiral emission or decoherence-free atomic interactions. Here, we show how to create giant atoms in dynamical state-dependent optical lattices, which offers the possibility of  coupling them to structured baths in arbitrary dimensions. This opens up new avenues to a variety of phenomena and opportunities for quantum simulation. In particular, we show how to engineer unconventional radiation patterns, like multi-directional chiral emission, as well as collective interactions that can be used to simulate non-equilibrium many-body dynamics with no analogue in other setups. Besides, the recipes we provide to harness giant atoms in high dimensions can be exported to other platforms where such non-local couplings can be engineered.
\end{abstract}

\maketitle

The design and exploration of novel forms of light-matter interaction have been a driving force in quantum optics triggering both fundamental and technological advances. A paradigmatic example of this was the observation that atomic lifetimes renormalize within cavities~\cite{purcell46a}, which opened the field of cavity QED~\cite{haroche89a,miller05a}. This seemingly simple light-matter coupling leaded to many other fundamental discoveries, such as the creation of mixed light-matter particles (polaritons), and applications, e.g., in quantum information~\cite{kimble08a}. Another timely example is the interaction of (natural or artificial) emitters with the structured propagating photons (or matter-waves) which appear in nanophotonics structures~\cite{vetsch10a,thompson13a,goban13a,beguin14a,lodahl15a,sipahigil16a,corzo16a,sorensen16a,solano17a,chang18a}, circuits~\cite{liu17a,sundaresan18a,mirhosseini18a}, or state-dependent optical lattices~\cite{devega08a,navarretebenlloch11a,ramos16a,vermersch16a,krinner18a}. In these systems, the bath displays structured energy dispersions, leading to a plethora of effects absent in other environments. On the fundamental level, they generate non-exponential relaxations~\cite{john94a,tong10a,longo10a,garmon13a,redchenko14a,lombardo14a,sanchezburillo17a}, whereas in the more applied perspective they lead to the emergence of bound states outside~\cite{bykov75a,john90a,kurizki90a,tanaka06a,calajo16a,shi16a} or in the continuum~\cite{facchi16a,galve17a,gonzaleztudela17a,gonzaleztudela17b,galve18a,asenjogarcia17a,shahmoon17a,glaetzle17a,perczel17a,albrecht18a}, which can be harnessed for (out-of) equilibrium quantum simulation~\cite{douglas15a,gonzaleztudela15c,shahmoon16a,gonzaleztudela18c}.

In all these setups the emitters are typically much smaller than their associated wavelength, leading to inherently local light-matter couplings. This picture, however, has been recently challenged with the design of the so-called "giant-atoms", which are emitters coupled to several points of SAW waveguides~\cite{gustafsson14a,manenti17a,noguchi17a,bolgar17a,moores18a} or transmission lines~\cite{ciani17a} separated beyond their characteristic wavelength. These giant atoms represent another paradigm change in quantum optics since the coupling to different bath positions induces strong interference effects which can be exploited for applications~\cite{kockum14a,ramos16a,vermersch16a,guo17a,kockum18a}. For instance, when coupled to one-dimensional baths they lead to decoherence-free atomic interactions~\cite{kockum18a}, or to chiral emission~\cite{ramos16a,vermersch16a} without exploiting polarization, something impossible to realize with "small" emitters. Exporting this paradigm to higher dimensional baths, where, for example, quantum simulation will show its full power, is a desirable, but challenging, goal. On the one hand, to our knowledge there is still no implementation to do so, since wiring up high-dimensional circuits becomes complicated. On the other hand, even if achieved, it is not obvious how to harness giant atoms when coupled to high dimensional baths. The reason is that the resonant photons mediating the interactions, defined by the isofrequencies of $\omega(\kk)$ at the emitters frequency, are contours (or surfaces) in the $\kk$-space, instead of points, making perfect interference more difficult.
\begin{figure}[tb]
	\centering
	\includegraphics[width=0.5\textwidth]{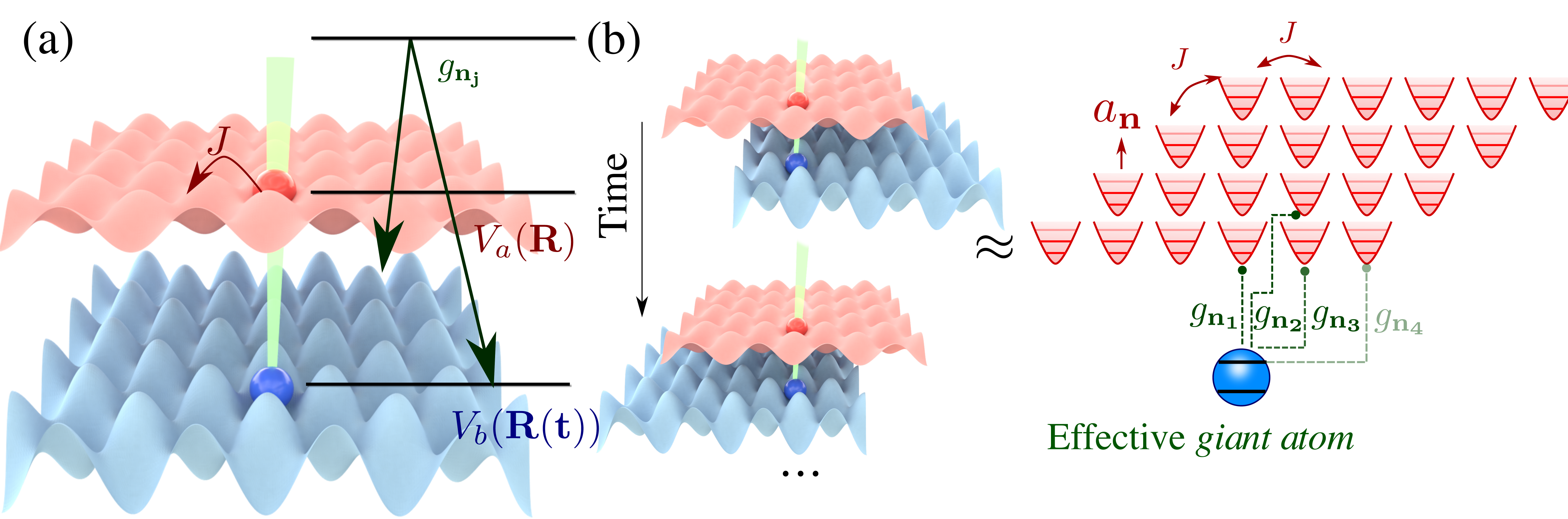}
	\caption{(a) State-dependent optical lattice scheme to simulate quantum optical phenomena: one deep lattice $V_b(\RR_b)$ (blue) traps the atomic state that mimics the QE behaviour, whereas a shallower one,  $V_a(\RR_a)$, lets matter-wave propagation at rate $J$. The two atomic state can be connected through a local laser(s) or microwave field (green) with strength $\Omega_{\nn_j}$. The relative position between the lattices, and of the local laser can be dynamically tuned $\RR_b(t)$. (b) Pictorial representation on how the effective giant atom couplings emerge from the stroboscopic movement between the lattices.}
	\label{fig:1}
\end{figure}

In this manuscript, we address both issues showing: i) A proposal to engineer effective giant atoms coupling to baths with high dimensions. We use ultra-cold atoms in dynamical state-dependent optical lattices~\cite{devega08a,navarretebenlloch11a,krinner18a} (see Fig.~\ref{fig:1}), such that by moving the relative position between the potentials~\cite{jaksch99b,sorensen99a,jane03a} fast enough, the giant emitter couples effectively to several bath positions. ii) A way to harness them to observe phenomena with no analogue in other setups by coupling them to structured photonic reservoirs with a Van-Hove singularity~\cite{gonzaleztudela17a,gonzaleztudela17b,galve17a,gonzaleztudela18c}. In particular, we show how giant quantum emitters (QEs) can modify the non-Markovian nature of the dynamics, and lead to unconventional emission patterns, e.g., chiral emission in one or several directions, which translate into unconventional collective QE interactions when several of them couple to the bath. Even though we make the discussion of (i-ii) together along the manuscript, the recipes that we provide for (ii) can be exported to other implementations where such couplings can be engineered.

\begin{figure}[tb]
	\centering
	\includegraphics[width=0.8\columnwidth]{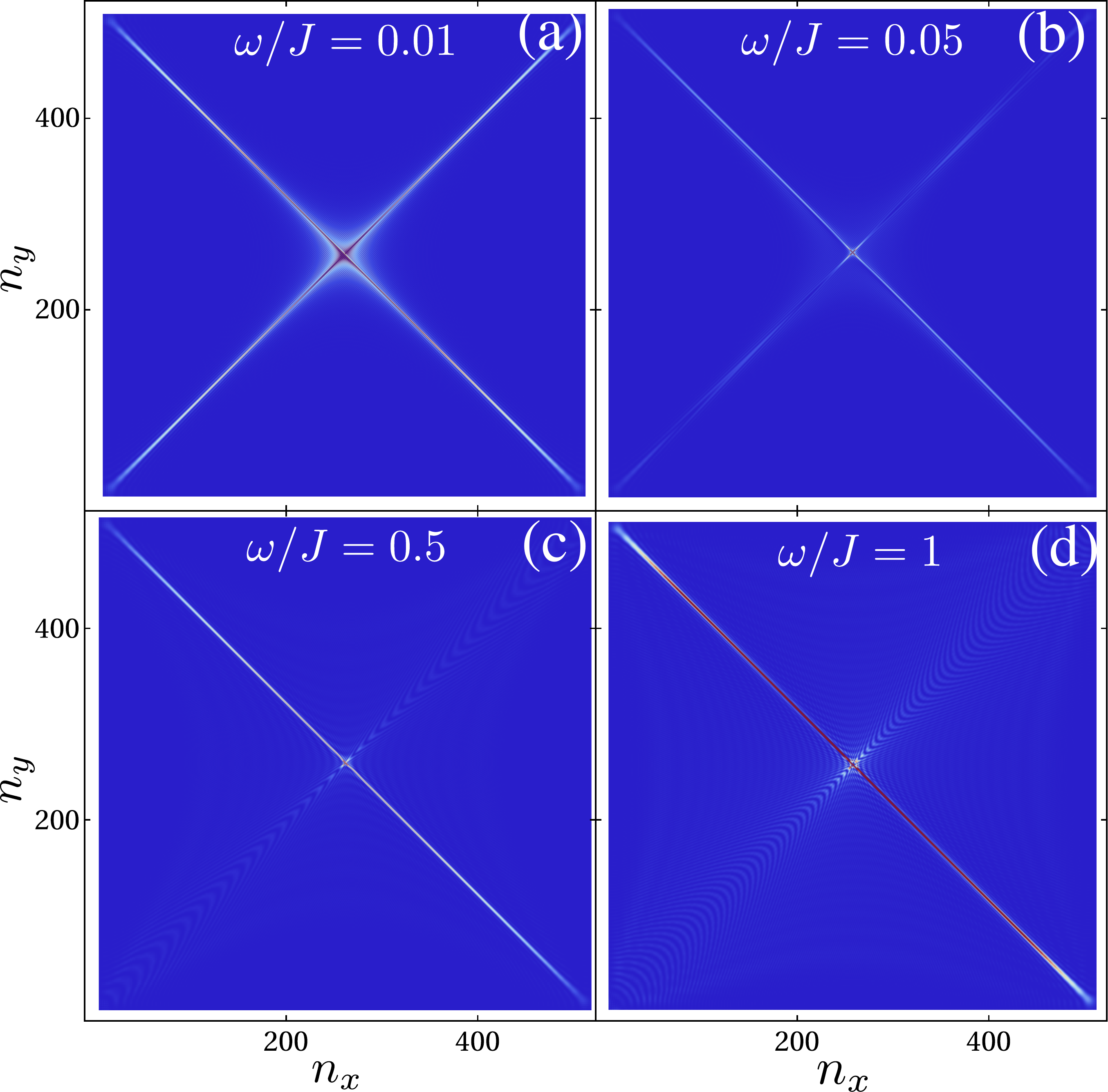}
	\caption{Bath population at a time $t J=N/4$  after the desexcitation of a single QE that moves between two lattice sites at positions $(0,0)$, $(1,1)$, such that $g_{\nn_1 [\nn_2]}=g\cos^2(\omega t/2) [\sin^2(\omega t/2)]$, with $g=0.1J$ and $\omega$ as depicted in the legend. Bath linear size is $N=512$. }
	\label{fig:2}
\end{figure}

Let us first recall how to obtain the standard quantum optical Hamiltonian with ultra-cold atoms~\cite{devega08a,navarretebenlloch11a,krinner18a}, see Fig.~\ref{fig:1}(a): one needs an atom with two states $a/b$ subject to different potentials $V_{a/b}(\RR)$, whose dimensionality can be optically controlled~\cite{bloch08a}. The $b$-atoms are trapped in a deep potential such that they mostly localize within a lattice site, and in the strongly interacting regime, which means that there will be at most one $b$-excitation per lattice site such that their excitations can be represented by spin operators $\sigma_{ge}^{\nn_j}$, with $\sigma_{\alpha\beta}^\nn=\ket{\alpha}_\nn \bra{\beta}$. On the contrary, when the atoms are in the $a$-state, they can hop to their nearest neighbours at a rate $J$ without interactions, mimicking photon propagation. Besides, one needs an extra field that transfers the $b$ excitations into $a$ ones (and viceversa), which can be obtained via a Raman or microwave transition~\cite{rubio18a,krinner18a} (or a direct one in the case of Alkaline-Earth atoms~\cite{daley08a,snigirev17a,riegger18a}). Let us denote as $\Omega_{\nn_j}$ the $a$-$b$ coupling at site $\nn_{j}$, which can be controlled in both magnitude and phase though the lasers. As derived in Refs.~\cite{devega08a,navarretebenlloch11a}, the Hamiltonian describing the dynamics of the excitations of the $a$ and $b$ atoms mimics the one standard light-matter interactions, that is, $H=H_S+H_B+H_\mathrm{int}$, where:
\begin{align}
H_S&=\omega_e \sigma_{ee}\,,\,\,
H_B=\sum_\kk \omega(\kk) a^\dagger_\kk a_\kk\,\label{eq:HSHB},\\
H_\mathrm{int}&=\left(\Omega_{\nn_e} a^\dagger_{\nn_e} \sigma_{ge}+\hc\right)\,,\label{eq:Hint}
\end{align}
where for illustration we restrict to a single QE, dropping the superindex in $\sigma_{ge}$.  The $a_\kk (a^\dagger_\kk)$ are the annihilation (creation) operator of a matter-wave excitation with momentum $\kk$, whose energy dispersion $\omega(\kk)$ is controlled by the geometry of $V_a(\RR)$. The QE is in the strong confinement limit such that its coupling will be local like with optical photons~\footnote{When the $b$ optical potential is weaker, its atomic wavefunction delocalizes allowing for a non-local coupling, which however does not help to obtain the desired behaviour.}. 

To effectively transform this local coupling into a non-local one among $\{\nn_\alpha\}_{\alpha=1}^{N_p}$ positions, one can dynamically move the relative position between the $V_{a/b}(\RR)$ potentials in a periodic fashion, e.g., changing the relative phase between the lasers creating the potentials~\cite{bloch08a}. If the movement is adiabatic, that is $|\dot{\RR}(t)|\ll d  \omega_{\mathrm{t}}$ for all $t$, where $d$ ground state size, and $\omega_{\mathrm{t}}$ the trap frequency~\cite{jaksch99b,sorensen99a,jane03a}, the atoms remain in their motional ground state and can still be described by a Hamiltonian as in Eqs.~\ref{eq:HSHB}-\ref{eq:Hint} but with time-dependent parameters. For example, assuming that the simulated QE probes the $\{\nn_\alpha\}_{\alpha=1}^{N_p}$ positions and that the laser parameter change as needed in each position, the Hamiltonian will now read:
\begin{align}
H_\intt\rightarrow H_{\intt,\mathrm{mov}}(t)=\sum_{\alpha=1}^{N_p} \left(\Omega_{\nn_\alpha}(t)a_{\nn_\alpha}^\dagger \sigma_{ge}+\hc\right)\,.
\end{align}

Now, to formally derive how the desired non-local couplings emerge using Floquet analysis, we consider that QE moves periodically along $N_p$ positions with period $T$ (and frequency $\omega=2\pi/T$), probing each position during a constant time interval $T/N_p$ with coupling strength $g_{\nn_{\alpha}}$~\footnote{In practice the transition from one position to the other will be a smooth function, which should satisfy the adiabaticity condition at any time.}. With that assumption, we can apply Floquet theory~\cite{goldman14a} to obtain an effective Hamiltonian description in the high-frequency limit. To the lowest order, it corresponds to the non-local light-matter couplings that we want to obtain (see Sup.~Material~\cite{SupMat}):
\begin{align}
 H_{\intt,\mathrm{eff}}\approx \sum_{\alpha=1}^{N_p} \left(\frac{g_{\nn_\alpha}}{N_p}a_{\nn_\alpha}^\dagger \sigma_{ge}+\hc\right)\,,\label{eq:hameff}
\end{align}
where $g_{\nn_\alpha}/N_p$ is the time average of $\Omega_{\nn_\alpha}(t)$. We can also calculate the next-order term contribution which is of order $\sim 4|g_\mathrm{max}|^2 N_p^2\zeta[3]/(\pi^2\omega)\ll \mathrm{max}|g_{\nn_\alpha}|$ for our situations of interest. Summing up, to obtain the desired behaviour the periodic movement has to be slow enough to stay within the lowest band of the tight-binding Hamiltonians of Eqs.~\ref{eq:HSHB}-\ref{eq:Hint}, but fast compared to the induced QE timescales, such that it effectively couples to several positions, i.e.,  $\omega_{\mathrm{t}}\gg \omega (L/d) \gg \mathrm{max}|g_{\nn_\alpha}|$ (assuming a constant speed over the distance $L$ that we displace the potentials). Since the couplings are tuneable and they can always be made small, the lower bound of these inequalities will be ultimately provided by the decoherence rate $\Gamma^*$ of the setup, which should be smaller than the simulated parameters. To provide some estimation, we can first take the recent realization of our proposed setup~\cite{krinner18a}, where two hyperfine $^{87}$Rb levels were used to engineer the optical potentials, $\ket{a/b}=\ket{F=1/2,m_F=-1/0}$, with trap-depths of the order $\omega_t\sim 2\pi\times 10$ kHz, and typical decoherence rates $\sim 10-100$~Hz. Another possibility is to use the ground/excited metastable state in Alkaline-Earth atoms~(see Ref.~\cite{daley08a} for a concrete proposal with Strontium). This platform shows similar $\omega_t$, but decoherence can be substantially decreased since it will be mostly determined by the excited state lifetime which can be $\Gamma^*/(2\pi)\lesssim 0.01$~Hz, thus leaving several orders of magnitude to adiabaticaly move the lattice. 

Let us now show how to exploit giant QEs coupled to higher dimensional baths to obtain phenomena with no analogue in other setups. In particular, we illustrate it by studying the spontaneous decay of an excited QE coupled to a two-dimensional bath with  $\omega(\kk)=\omega_a-2J\left[\cos(k_x)+\cos(k_y)\right]$.
When the QE interacts locally in space with frequency $\omega_e=\omega_a$, it couples equally to all the resonant $\kk$'s defined by: $k_x\pm k_y =\pm (\mp) \pi$. This contour, which includes points with zero group velocity ($\vv_g(0,\pm \pi)=\vv_g(\pm \pi,0)=(0,0)$) responsible of a Van-Hove singularity in the density of states~\cite{vanhove53a}, leads to two remarkable effects in the QE spontaneous decay~\cite{gonzaleztudela17a,gonzaleztudela17b,galve17a}: i) its emission pattern is highly anisotropic, as shown in Fig.~\ref{fig:2}(a), emitting mostly in four directions with some diffraction due to the inhomogeneous group velocity of the wavepacket; ii) its dynamics is intrinsically non-Markovian due to divergence of the density of states at this frequency~\cite{gonzaleztudela17a,gonzaleztudela17b}. Now, we will show how building up on this behaviour, giant QEs can lead to very flexible and unusual emission patterns and interactions.

\begin{figure}[tb]
	\centering
	\includegraphics[width=0.89\columnwidth]{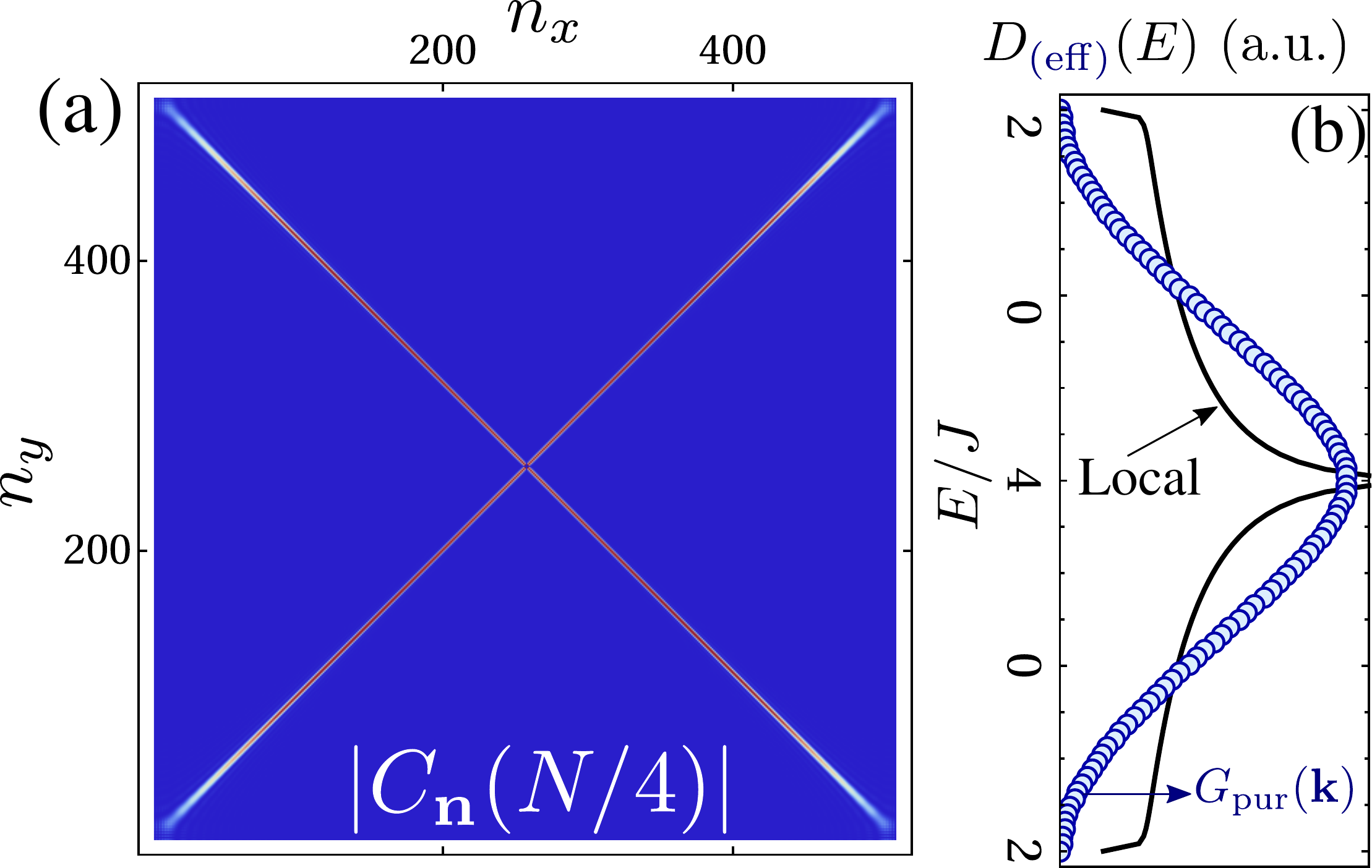}
	\caption{(a) Bath probability amplitude at a time $t J=N/4$ for a single giant QE coupled with $G_\mathrm{pur}(\kk)$, $g=0.1J$ and bath linear size is $N=512$. (b) Comparison of the density of states of the structured bath (black), with the effective spectral density which includes the $\kk$ dependence of $G_\mathrm{pur}(\kk)$, i.e., $D_\mathrm{eff}(E)=\sum_\kk |G_\mathrm{pur}(\kk)|^2\delta(E-\omega(\kk))$.   }
	\label{fig:3}
\end{figure}

\emph{Quasi-1D emission.} 
First, we show how to cancel the emission in one of the diagonals of Fig.~\ref{fig:2}(a) by coupling to two lattice sites $\nn_{1/2}=(0,0)/(1,1)$. To numerically show how the Floquet averaged Hamiltonian $H_{\intt,\mathrm{eff}}$ emerges, we assume that the movement between the lattices is such that $\Omega_{\nn_1}(t)=g\cos^2(\omega t/2) $, $\Omega_{\nn_2}(t)=g\sin^2(\omega t/2)$, and solve the dynamics using $H_{\mathrm{int},\mathrm{mov}}(t)$. In Fig.~\ref{fig:2}(b-d) we plot the bath population in real space after a time $tJ=N/4$ using $g=0.1J$, and for several $\omega$'s. As expected, for $\omega\ll g$, the emission occurs in four directions as if the QE was locally coupled. However, as $\omega$  increases, the interference between the bath emission in two different points occurs, until it cancels the emission in one of the diagonals. This behaviour can be understood from the asymptotic bath state in the perturbative limit~\cite{gonzaleztudela17b}:
\begin{equation}
\label{eq:batht}
C_\kk(t\rightarrow \infty)\propto \frac{G(\kk)e^{-i\omega(\kk)t}}{\omega(\kk)-\omega_e+i\Gamma_M/2}\,,
\end{equation}
where $\Gamma_M$ is the Markovian decay rate, and $G(\kk)$ is the effective light-matter coupling between the emitter and the $\kk$-modes, $H_{\intt,\mathrm{eff}}=\sum_\kk \left(G(\kk)a^\dagger_\kk \sigma_{gs}+\hc\right)$, which reads:
\begin{align}
G(\kk)=\frac{1}{N_p}\sum_{\alpha=1}^{N_p} g_{\nn_\alpha} e^{- i\kk\cdot \nn_\alpha}\,.
\end{align}

In this case $G_\mathrm{1D}(\kk)\propto 1+e^{-i(k_x+k_y)}$, which satisfies $G_\mathrm{1D}(k_x,\pm \pi-k_x)\equiv 0$. Thus, the giant QE is effectively uncoupled from the $\kk$-modes responsible of the forward/backward direction in the diagonal where the giant QE is coupled to, and does not decay into them. After having numerically seen how $H_{\intt,\mathrm{eff}}$ emerges from $H_{\mathrm{int},\mathrm{mov}}(t)$ for this example, from now on we use $H_{\intt,\mathrm{eff}}$ to analyze the dynamics.

\emph{Trapped emission.} 
Let us now consider that the QE moves around four positions, i.e., $(\pm 1,0),(0,\pm 1)$. The effective $\kk$-coupling will be:
 $G_\mathrm{trap}(\kk)=g\left( e^{i k_x}+e^{-i k_x}+e^{i k_y}+e^{-i k_y}\right)/4$, which cancels the coupling among the four resonant $\kk$-lines. Thus, the giant QE will not decay, while keeping some the photon population trapped between the four positions (not shown). As in the 1D counterpart~\cite{kockum18a}, these confined photons will mediate coherent interactions between these decoherence-free QEs.

\emph{Filtering non-Markovian emission.}  
Another feature that can be achieved by coupling to few lattice sites is the effective decoupling from zero-group velocity terms occurring at $\kk=(0,\pm \pi)$ and $(\pm \pi,0)$. For that, we can couple the QE to the positions $(\pm 1,\pm 1), (\pm 1,\mp 1)$,  with an alternating $\pm 1$ phase, such that $G_\mathrm{pur}(\kk)=g\sin(k_1)\sin(k_2)$. This has two consequences: first, the QE shows a more homogeneous directional emission, as observed in Fig.~\ref{fig:3}(a). Second, it smoothens the effective spectral density probed by the QE, as plotted in Fig.~\ref{fig:3}(b), making its dynamics more Markovian. Thus, giant QEs provide a way of decoupling directional emission from non-Markovian dynamics in Van-Hove singularities.

\begin{figure}[tb]
	\centering
	\includegraphics[width=0.99\columnwidth]{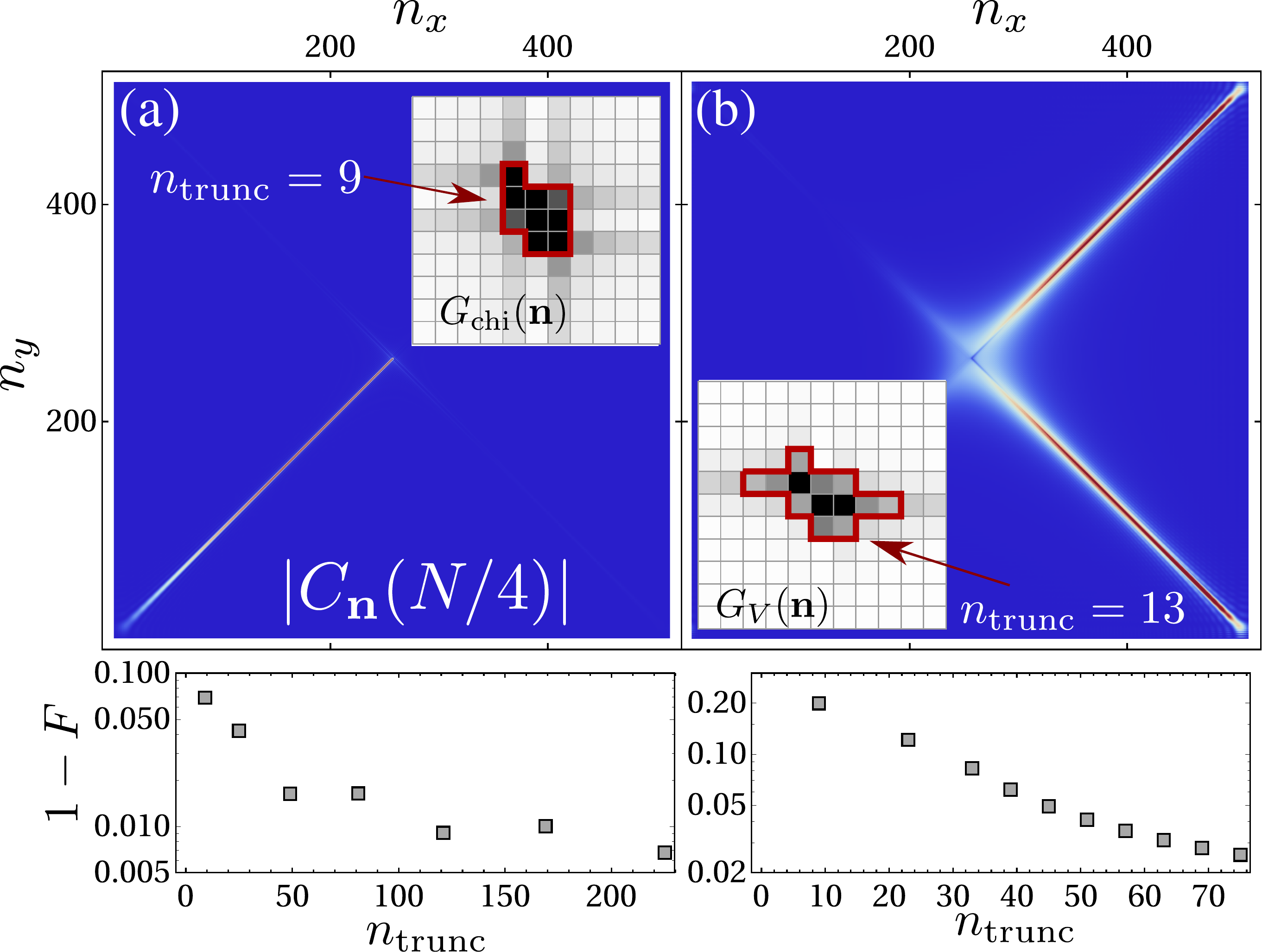}
	\caption{ (a-b) Bath probability amplitude at a time $t J=N/4$ for a single giant QE coupled with $G_\mathrm{trunc}(\nn;n_\mathrm{tr})$, respectively, $g=0.1J$ and bath linear size is $N=512$. Inset: Corresponding spatial coupling profile $G(\nn)$ using Eq.~\ref{eq:recipett}. In red the truncation we use to plot the figure. (c-d) $1-F$, where $F$ is the fraction of the emission into the desired directions for the parameters of panel (a-b), as a function of the number of terms in the sum  $G_\mathrm{trunc}(\nn;n_\mathrm{tr})$.}
	\label{fig:4}
\end{figure}

\emph{Reverse design: chiral and V-type emission.}
In the previous examples it was possible to guess the spatial couplings required to obtain the desired behaviour. An alternative approach consists of first guessing the $G(\kk)$ required to obtain a given behaviour, and then Fourier transforming it to get the spatial dependent couplings, that is
\begin{align}
\label{eq:recipett}
G(\nn)=\frac{1}{N^2}\sum_{\kk}G(\kk) e^{-i\kk\cdot\nn}\,.
\end{align}

For example, let us imagine we want to obtain perfect chiral emission in one or two orthogonal directions out of the four appearing with local couplings. It is easy to see that:
\begin{align}
\label{eq:chiralf}
G_\mathrm{chi}(\kk)&\propto \cos\left(\frac{k_1-k_2}{2}\right)\left[1+\sin\left(\frac{k_1+k_2}{2}\right)\right]\,,\\
G_V(\kk)&\propto \left[1-\sin\left(\frac{k_1-k_2}{2}\right)\right]\left[1-\sin\left(\frac{k_1+k_2}{2}\right)\right]\,.\label{eq:V}
\end{align}
cancels the coupling to the light emitted in three (or two) of the four directions, respectively. Then, using Eq.~\ref{eq:recipett} we obtain the spatial profile of the couplings whose absolute value $|G(\nn)|$ is plotted in the inset of Figs.~\ref{fig:4}(a-b). The coupling spatial pattern is more intricate than in the previous situations because it requires adding complex phases (not shown), and involve the coupling to many lattice sites. Since this will be experimentally challenging, one needs to adopt a truncation strategy in which one approximates the sum by a finite number $n_\mathrm{tr}$ of terms, $G(\nn)\approx G_\mathrm{trunc}(\nn;n_\mathrm{tr})$. This is what we do in Figs.~\ref{fig:4}(a-b), where we observe that even for a small $n_\mathrm{tr}$, the QE emits approximately with the desired behaviour. Finally, in Fig.~\ref{fig:4}(c-d) we show how increasing $n_\mathrm{tr}$, the light collimated in the desired directions can go close to 100 \%. 

\emph{Interactions.}
Let us finally point how these unconventional emission patterns will translate into exotic QE interactions when $N_e$ QEs are coupled to the bath. For simplicity, let us assume that each QE has a $\kk$-dependent coupling $G_j(\kk)=G(\kk)e^{-i\kk\cdot \nn_j}$, where $e^{-i\kk\cdot \nn_j}$ is a global phase factor which indicates the giant QE central position ($\nn_j$), and $G(\kk)$ is a common $\kk$ dependent coupling defined by the non-local couplings around the position $\nn_j$. Then, if we trace out the bath degrees of freedom under the Born-Markov approximation, the QE reduced density matrix ($\rho)$ dynamics is governed by~\cite{gardiner_book00a}: $\partial_t\rho=i[\rho,H_S+\sum_{i,j}J_{ij}\sigma_{eg}^i\sigma_{ge}^j]+\sum_{i,j}\gamma_{ij}/2(2\sigma_{ge}^i\rho\sigma_{eg}^j-\sigma_{eg}^j\sigma_{ge}^i\rho-\rho\sigma_{eg}^j\sigma_{ge}^i)$. The collective interactions $J_{i,j},\gamma_{ij}$ are:
\begin{align}
\label{eq:int}
\frac{\gamma_{ij}}{2}+i J_{ij}=\frac{1}{N}\sum_\kk\frac{|G(\kk)|^2 }{\omega_e-\omega(\kk)+i0^+} e^{i\kk\cdot(\nn_i-\nn_j)}\,,
\end{align}
whose integrand is directly connected with the asymptotic emission pattern described in Eq.~\ref{eq:batht}. This tells us, for example, that using the couplings $G_{1D}(\kk)$ or $G_{\mathrm{chi}}(\kk)$ we will be able to simulate standard or chiral~\cite{lodahl17a} waveguide QED couplings in two-dimensional baths, as well as other QE interactions with no counterpart in other setups, i.e., V-type collective decays.

\emph{Conclusions.}
Summing up, we propose a method to engineer effective non-local light-matter couplings using ultra-cold atoms in dynamical state-dependent optical lattices. Controlling the confinement and relative position of two optical potentials, one can simulate giant atoms coupled to structured photonic baths in one, two and three dimensions. Irrespective of the implementation,  we also numerically illustrate the potential of giant emitters to yield unconventional quantum optical behaviour when coupled to a two-dimensional structured bath. In particular, we exploit the interplay between the structured energy dispersion and non-local couplings to obtain exotic emission patterns and collective dissipative interactions. These recipes can be immediately adapted to other platforms where such non-local couplings can be engineered, or to higher dimensions~\cite{gonzaleztudela18d,SupMat}.

Beyond the fundamental interest of the phenomena explored along the manuscript, there are many possible follow-up applications. From the quantum simulation perspective, giant atoms provide a very flexible playground to probe equilibirum~\cite{douglas15a,gonzaleztudela15c} and non-equilibrium many-body physics~\cite{ramos14a,pichler15a} with no analogue in other setups. Besides one can increase their tunability exploiting the interplay with the polarization degree of freedom~\cite{perczel18a,perczel18b,yu18a}, or through additional bath engineering~\cite{gonzaleztudela18f}. Other possibilities, if one is able to engineer it with optical photons, is to exploit the multi-directional chiral emission to transfer simultaneously quantum information into several nodes, or for generating high-dimensional photonic entangled states~\cite{pichler17a}, which can be used for fault-tolerant measurement based quantum computation~\cite{briegel09a}.

\section{Acknowledgements}

JIC acknowledges the ERC Advanced Grant QENOCOBA under the EU Horizon 2020 program (grant agreement 742102).  C.S.M. is supported by the Marie Sklodowska-Curie Fellowship QUSON (Project No. 752180). AGT acknowledges very useful discussions with J.~Kn\"orzer, and the critical reading of the manuscript of T.~Ramos.


\bibliography{Sci,books}

\newpage
\begin{widetext}
	\widetext
	\onecolumngrid
	\begin{center}
		\textbf{\large Supplemental Material: Tailoring quantum optical phenomena shaking state-dependent optical lattices}
	\end{center}
	\vspace{\columnsep}
	\vspace{\columnsep}
	
	\twocolumngrid
\end{widetext}
\setcounter{equation}{0}
\setcounter{figure}{0}
\setcounter{section}{0}
\makeatletter

\renewcommand{\thefigure}{SM\arabic{figure}}
\renewcommand{\thesection}{SM\arabic{section}}  
\renewcommand{\theequation}{SM\arabic{equation}}  

In this Supplementary Material, we provide more details on: i) the derivation of the time-averaged Hamiltonian in the high-frequency limit; ii) how to calculate the radiation patterns from spontaneous emission; iii) experimental setup and feasibility analysis; iv) how to extend some of the phenomena predicted in the manuscript to three-dimensional systems.

\section{Deriving the Floquet Hamiltonian\label{secSM:floquet}}

The starting point of the derivation is the time-dependent interaction Hamiltonian of Eq. 4 of the main text:
\begin{align}
H_{\mathrm{int},\mathrm{mov}}(t)=\sum^{N_p}_{\alpha=1} \left(\Omega_{\nn,\alpha}(t)a_{\nn_\alpha}^\dagger \sigma_{ge}+\hc\right)\,.
\end{align}

To do the Floquet derivation, we assume that we move the lattices such that the QE excitation probes $N_p$ positions, each of them during a time $T/N_p$, such that the global period of the movement is $\omega=2\pi/T$. Thus, $\Omega_{\nn,\alpha}(t)=g_{\nn,\alpha}f_\alpha(t)$, where:
\begin{align}
f_\alpha(t)&=1,\,(\alpha-1)T/N_p<t<\alpha T/N_p\,,
\end{align}
or $0$ otherwise. We are aware that in a practical situation the transition from one lattice site to the other will be smooth and, in fact, must satisfy the adiabaticity condition at any time~\cite{jaksch99b,sorensen99a,jane03a}. However, the step function $f_\alpha(t)$ will allow us to capture analytically the most relevant features of the effective dynamics without worrying about the particular details of $\Omega_{\nn,\alpha}(t)$. 

The key point is that step functions can be easily expanded in their Fourier components  as follows~\cite{goldman14a}:
\begin{align}
f_\alpha(t)&=\frac{1}{N_p}+\sum_{j}C_{j,\alpha} e^{i j \omega t}\,,\\
C_{j,\alpha}&=\frac{1}{2\pi i j}e^{-i 2\pi\alpha j/N_p}\left(e^{i2\pi j/N_p}-1\right)
\end{align}

Using this expansion, the Hamiltonian $H_{\mathrm{int},\mathrm{mov}}(t)$ can be separated into a time-independent part, $V^{(0)}$, which contains a the time-averaged interaction of $H_{\mathrm{int},\mathrm{mov}}(t)$,
\begin{align}
V^{(0)}=\frac{1}{N_p}\sum_{\alpha}\left(g_{\nn_\alpha}a^\dagger_{\nn_\alpha}\sigma_{ge}+\hc\right)\,,
\end{align}
that is the part we want to obtain, plus all the periodic modulation introduced by the harmonics:
\begin{align}
H_{\mathrm{int},\mathrm{mov}}(t)-V^{(0)}=\sum_{j\neq 0}V^{(j)} e^{ij\omega t}\,,\\
V^{(j)}=\sum_{\alpha}\left(C_{j,\alpha} g_{\nn_\alpha}a^\dagger_{\nn_\alpha}\sigma_{ge}+\hc\right)\,,
\end{align}

In the high-frequency limit, an effective time-independent Hamiltonian can be derived~\cite{goldman14a}, which to first order in $1/\omega$ reads:
\begin{align}
H_{\mathrm{int},\mathrm{eff}}\approx V^{(0)}+\frac{1}{\omega}\sum_{j>0}\frac{[V^{(j)},V^{(-j)}]}{j}\,
\end{align}

With it, we can calculate explicitly the first order correction to the time averaged Hamiltonian $V^{(0)}$:
\begin{widetext}
\begin{align*}
H^{(1)}_\mathrm{eff}&=\frac{1}{\omega}\sum_{j>0}\frac{[V^{(j)},V^{(-j)}]}{j}=\sum_{\alpha,\beta}^{N_p}\sum_{j=1}^{\infty}\left[\frac{4  ig_{\nn_\alpha}g^*_{\nn_\beta} }{\pi^2 j^3\omega}\sin^2\left(\frac{j\pi}{N_p}\right)\sin\left(\frac{2\pi(\beta-\alpha)j}{N_p}\right)a^\dagger_{\nn_\alpha}a_{\nn_\beta}\sigma_z\right]\,
\end{align*}
\end{widetext}
with $\sigma_z=(\sigma_{ee}-\sigma_{gg})/2$. Since we typically restrict to situations where the number of excitations, $\hat{N}=\sum_\nn a_\nn^\dagger a_\nn +\sigma_{ee}$, is conserved, and in the single-excitation regime, the norm of this operator can be upper-bounded by:
	\begin{align}
	||H^{(1)}_\mathrm{eff}||<\frac{4g^2 N_p^2}{\pi^2\omega}\sum_{j=1}^{\infty}\frac{1 }{j^3}=\frac{4g^2 N_p^2}{\pi^2\omega}\zeta[3]	\end{align}
where $g=\mathrm{max}\{|g_\nn|\}$. It must be noted that when considering the full dynamics with $H_S+H_B$, the density of states will also enter into play in the discussion, i.e., suppressing (or enhancing) the contributions of the different sidebands at frequencies $j\omega$. In the examples considered along the text, since the density of states is peaked around $\omega_a$, the sideband contributions are suppressed compared to the time-averaged component. The opposite behaviour (enhancement of sidebands) can also be used an extra degree of freedom to design more exotic quantum optical phenomena beyond the time averaged terms of $H_{\mathrm{int},\mathrm{mov}}(t)$.

 \section{Calculating the emission patterns \label{secSM:emission}}
 
 The global Hamiltonian of the system: $H=H_S+H_B+H_\intt$ conserves the number of excitations: $\hat{N}=\sum_{j}\sigma_{ee}^j+\sum_\kk a^\dagger_\kk a_\kk$, no matter whether $H_\intt$ is time-dependent or not. Thus, if we consider a single QE initially excited as the initial state: $\ket{\Psi(0)}=\ket{e}\otimes\ket{\mathrm{vac}}$, the global state at any time can be written as:
 \begin{align}
 \ket{\Psi(t)}=\left[C_e(t)\sigma_{eg}+\sum_\nn C_\nn(t) a_\nn^\dagger\right]\ket{g}\otimes\ket{\mathrm{vac}}\,,
 \end{align}
 where the coefficients can be always obtained by numerically solving $i\frac{\ket{\Psi(t)}}{dt}=H(t)\ket{\Psi(t)}$. This is how we obtain the $C_\nn(t)$ plotted in the Figs.~2-4 of the main text. Moreover, by Fourier transforming $C_\nn(t)$ we can obtain the wavefunction in momentum space:
 \begin{align}
\label{eq:recipet}
C_\kk(t)=\frac{1}{N^2}\sum_{\nn}C_\nn(t) e^{-i\kk\cdot \nn}\,.
\end{align}
 
 With $C_\kk(t)$ it is easy to define the fraction of light emitted in each of the four directions of Fig.~2(a) at any time:
 \begin{align}
 F_{1}(t)&=\frac{\sum_{k_x>0,k_y>0}|C_{\kk}(t)|^2}{C_{\kk}(t)}\,,\\
 F_{2}(t)&=\frac{\sum_{k_x<0,k_y>0}|C_{\kk}(t)|^2}{C_{\kk}(t)}\,,\\
 F_{3}(t)&=\frac{\sum_{k_x<0,k_y<0}|C_{\kk}(t)|^2}{C_{\kk}(t)}\,,\\
 F_{4}(t)&=\frac{\sum_{k_x>0,k_y<0}|C_{\kk}(t)|^2}{C_{\kk}(t)}\,.
 \end{align}
 
 This is what we use to characterize the fraction of light emitted in one or two-directions in Fig.~4(c-d). For the chiral emission we plot $1-F_3(t)$, and for the $V$-shape emission $1-F_1(t)-F_4(t)$ at time $tJ=N/4$ for the parameters written in the caption.\\

 \section{Experimental considerations \label{secSM:experimental}}
 
 In this Section we give a more detailed explanation on the experimental setup that could be used to observe the phenomena predicted in the manuscript, and analyze the feasibility of our proposal using realistic experimental parameters.
 
 \subsection{Atomic level configuration}
 
\begin{figure}[!tb]
 	\centering
 	\includegraphics[width=0.99\columnwidth]{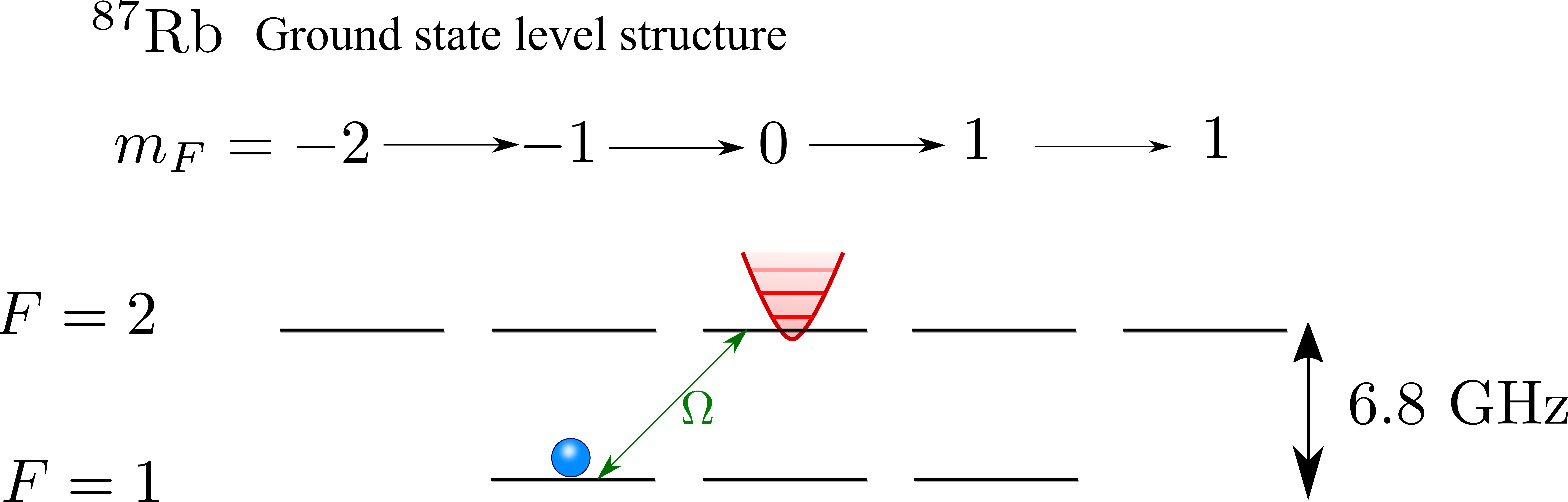}
 	\caption{Ground state level structure of $^{87}$~Rb with two hyperfine ground state manifolds ($F=1$ and $F=2$) with three and five states labeled by its $m_F$ quantum number, respectively. The two ground states manifolds are separated in energies by $6.8$~GHz. In Reference~\cite{krinner18a}, they use the states $\ket{F=1,m_F=-1}$ and $\ket{F=2,m_F=0}$, as the quantum emitter/bath states, respectively, and a microwave field, $\Omega$, to couple them.}
 	\label{figSM:1}
 \end{figure}

 One possibility consists in using Rubidium atoms as in the recent experiment by Krinner \emph{et al}~\cite{krinner18a}, where quantum optical phenomena was simulated for the first time using Rubidium matter-waves in state-dependent optical potentials. As schematically explained in Fig.~\ref{figSM:1}, in that experiment two states in the ground state manifold of $^{87}$Rb atoms are used to simulate the quantum emitter and bath. Let us review some of the parameters of that experiment:
 \begin{itemize}
 	\item They use the $\ket{F=1,m_F=-1}=\ket{b}$ and $\ket{F=2,m_F=0}=\ket{a}$, as the emitter/bath state, respectively, which are separated in energies by $6.8$~GHz. 
 	
 	\item They transfer the excitations directly from $a$ to $b$ using a microwave field with strength of the order of $\Omega/2\pi\sim 1$~kHZ.
 	
 	\item They are interested in observing one-dimensional band-edge physics, such that they enforce the two atomic states to live within one-dimensional tubes through a common radial confinement. The state-dependent optical potential along the other direction is generated with a $\sigma^{-}$-polarized laser beam with $\lambda=790$~nm. They choose that combination of polarization/wavelength such that the $a$ atom does not feel any potential along that direction, while the emitter-like state is strongly confined with a trap frequency $\omega_t/(2\pi)\sim 40$~KHz. However, as they mention in their Sup.~Material by either rotating the polarization and/or changing wavelength, they can also induce different trapping conditions for $a$. 
 	
 	\item The advantage of using Rb hyperfine states is that they have very long coherence times. Possible sources of decoherence such as thermal fluctuations or the scattering rates introduced by the trapping potentials are very well understood and under control in these setups. For example, the main source of these spin-dependent lattices will be the scattering rates introduced by the optical potential, which for that particular experiment, we estimate to be $\Gamma^*/(2\pi)\sim 10-100$ Hz (even though it was not explicitly mentioned in the paper). By using different atomic states, and/or wavelengths one could optimize these decoherence rates for the particular experiments we are considering.
\end{itemize}

A variation of this setup can be used to implement our ideas. One would require: i) extra laser fields to create the optical confinement in other directions depending on whether we want to simulate two or three dimensional baths; ii) a way of dynamically change the phase of the laser to displace the emitter-state optical potential. Besides, if want the bath state to create a fully independent tuneable optical potential for the bath state, one should add independent laser fields with other polarization/frequencies.

Another interesting possibility consists of using Alkali-Earth atoms~\cite{daley08a}, such as Ytterbium~\cite{riegger18a} or Strontium~\cite{snigirev17a}, to create such state-dependent optical lattices. These atoms are characterized by having optically excited mestable states, $^3P_{0/2}$ with very narrow linewidths which can be as small as $\Gamma_e/(2\pi)\sim 0.01$~Hz. This allows one to use these excited states to store excitations with very long coherence times. The advantage is that since the ground and excited states are separated by optical frequencies, one can engineer completely independent potentials for both states. For a particular realization of such independent state-dependent optical lattices with Strontium one can check Ref.~\cite{daley08a}, where it was also explained how to dynamically move the relative position between the two potentials, and how to transfer the excitations between the states.
 
 \subsection{Feasibility conditions}
 
 As discussed in the main text, the approximate set of inequalities to obtain the desired phenomena is lower and upper bounded by the decoherence rate and trap frequencies, respectively. Thus, the difference between these two magnitudes determine how much room we have to implement our proposal, for example, limiting how many bath positions the emitter can probe without breaking the adiabaticity condition. In the previous Section, we have shown that typical experimental parameters for, e.g., Alkali-Earth atoms, can be $\omega_t/2\pi\sim 10^4$~Hz, and $\Gamma^*/(2\pi)\sim 0.01$~Hz, such that one in principle has many orders of magnitude available to play with. In any case, we want to note that when implementing our ideas in a particular setup, a full analysis of all possible error sources and limitations should be performed to fully understand the limits of the experiment.\\

  \section{Giant atoms in three-dimensional baths \label{secSM:3Dbaths}}
 
 As we argue in the concluding paragraph of the main text, many of the phenomena and recipes that we illustrate for two-dimensional photonic baths can be exported to three-dimensional ones with an adequate choice of the bath and emitter-bath couplings. For example, as shown in Ref.~\cite{gonzaleztudela18d}, if the three-dimensional bath has a body-centered-cubic geometry, the energy dispersion is given by:
 \begin{align}
 \omega(\kk)=-2J\left[\cos(k_x)+\cos(k_y)+\cos(k_z)+\cos(k_x+k_y+k_z)\right]\,.
 \end{align}
 
 This energy dispersion leads to a Van-Hove singularity in the middle of the band, $\omega(\kk)=0$, which occurs for the planes $k_a\pm k_b=\pm \pi$, where $a,b$ is any combination of $x,y,z$. This leads to a highly directional emission pattern in eight lines, as compared to four lines in 2D, which we can exploit in combination with giant atoms in a similar fashion. For example,  
 \begin{itemize}
 	\item Coupling an emitter to two positions, e.g., $(0,0,0)$, $(1,1,0)$, the $\kk$-dependent coupling will read $G(\kk)\propto 1+e^{i(k_1+k_2)}$ that will vanish when $k_{1}+k_{2}=\pm \pi$, canceling the emission into the direction defined these planes. Using these tricks, or directly applying the reverse engineering that we explain in the main text, one can obtain the connectivity that one desires along these eight emitting lines defined by energy dispersion of the bath.
 	
 	\item The latter also includes the design of decoherence-free atoms in three-dimensions coupling the emitter to eight positions instead of four.
 	
 \end{itemize}
 
Apart from extending the control in Van-Hove singularity points, three-dimensional reservoirs also display other types of radiation patterns, such as emission in directional planes in cubic simple geometries~\cite{gonzaleztudela18f}. These could as well lead to interesting effects when combined with giant atoms. To avoid overloading the manuscript, we leave the detailed study of the three-dimensional scenario for a separate work.
 
\end{document}